\documentclass[10pt,a4paper,notitlepage]{iopart_noeqstarred}
\usepackage[numbers,sort&compress,comma]{natbib}
 
\usepackage{geometry}
\usepackage{amsmath,amssymb}
\usepackage{graphicx,color}
\usepackage{hyperref}

\definecolor{rltred}{rgb}{0.75,0,0}
\definecolor{rltgreen}{rgb}{0,0.5,0}
\hypersetup{colorlinks,hypertexnames=true,pdfhighlight=/I,linkcolor=rltred,citecolor=rltgreen}

\begin{document}

\twocolumn[ 

\title{Time-resolved photoemission by attosecond streaking: extraction of time information}

\author{S.~Nagele$^1$, R.~Pazourek$^1$, J.~Feist$^2$, K.~Doblhoff-Dier$^1$, C.~Lemell$^1$, K. T\H{o}k\'{e}si$^3$, J. Burgd\"orfer$^4$}
\address{$^1$ Institute for Theoretical Physics, Vienna University of Technology, 1040 Vienna, Austria, EU}
\address{$^2$ ITAMP, Harvard-Smithsonian Center for Astrophysics, Cambridge, Massachusetts 02138, USA}
\address{$^3$ Institute of Nuclear Research of the Hungarian Academy of Sciences (ATOMKI), 4001 Debrecen, Hungary, EU}
\ead{\mailto{stefan.nagele@tuwien.ac.at}, \mailto{renate.pazourek@tuwien.ac.at}}

\date{\today}

\begin{abstract}
Attosecond streaking of atomic photoemission holds the promise to provide unprecedented information on the release time of the photoelectron. We show that attosecond streaking phase shifts indeed contain timing (or spectral phase) information associated with the Eisenbud-Wigner-Smith time delay matrix of quantum scattering. However, this is only accessible if the influence of the streaking infrared (IR) field on the emission process is properly accounted for. The IR probe field can strongly modify the observed streaking phase shift. We show that the part of the phase shift (``time shift'') due to the interaction between the outgoing electron and the combined Coulomb and IR laser fields can be described classically. By contrast, the strong initial-state dependence of the streaking phase shift is only revealed through the solution of the time-dependent Schr\"odinger equation in its full dimensionality. We find a time delay between the hydrogenic $2s$ and $2p$ initial states in He$^+$ exceeding 20\,as for a wide range of IR intensities and XUV energies. 
\end{abstract}

\vspace{-8mm}
\vspace{5mm}

\thispagestyle{plain}
\maketitle

] 

\pagestyle{plain}

Attosecond streaking is one of the most spectacular applications within the emerging field of \emph{attoscience} \cite{HenKieSpi2001, DreHenKie2001, ItaQueYud2002, KieGouUib2004, YakBamScr2005, QueMaiIta2005, SanBenCal2006}. It has developed into a powerful tool to achieve time-resolution on the sub-100 attosecond time scale \cite{DreHenKie2002, GouUibKie2004, CavMueUph2007, SchFieKar2010}. Streaking is based on a variant of a pump-probe setting with an extreme ultraviolet (XUV) pulse of a few hundred attoseconds serving as pump and a phase-controlled few-cycle infrared (IR) pulse as probe. Electrons emitted in the presence of an IR field are accelerated to different final momenta and energies depending on the value of the vector potential at the release time. Thus, time information is mapped onto the energy axis in analogy to conventional streaking. First proof-of-principle implementations were the direct time-domain measurement of the life time of the Xe($4p^{-1}$) hole by Auger decay of $\simeq\!8$\,fs \cite{DreHenKie2002} complementing equivalent spectral information, as well as the measurement of the time-dependent electric field of the IR light wave \cite{GouUibKie2004}. More recent and more advanced applications addressed the delayed photoemission from core levels relative to conduction band states of a tungsten surface due to the increased pathlength and inelastic scattering processes \cite{CavMueUph2007}. Very recently, a time delay of atomic photoemission from the $2s$ shell relative to the $2p$ shell of Neon as small as $21$ attoseconds, i.e.\ less than one atomic time unit has been extracted \cite{SchFieKar2010}. This time scale is one order of magnitude shorter than the XUV ``pump'' pulse and two orders shorter than the oscillation period $T$ of the IR probe pulse. First theoretical model calculations \cite{SchFieKar2010,KheIva2010} could qualitatively account for a delay of the $2p$ relative to the $2s$ electron, falling, however, short by more than a factor of 3 to quantitatively reproduce the $\simeq\!20$\,as delay. 

As the resolution of the phase of the streaking field reaches such a small fraction of an optical cycle that the associated time resolution approaches the atomic time scale (a classical electron in the ground state of H travels 0.5\,\AA\ during $\simeq\!24.2$\,as), fundamental conceptual questions arise as to the physical information to be accessed. More generally, the question is posed which novel information can be extracted in the time domain which cannot be gained by high-resolution measurements in the spectral domain. For complex systems in the presence of incoherent processes (such as electron transport near surfaces \cite{LemSolTok2009}), time domain studies of relaxation and transport can obviously provide novel information on competing decoherent pathways not easily accessible in the spectral domain. For fully coherent quantum processes where time and spectral domain descriptions are, in principle, linked by a Fourier transform, the extraction of novel information is of considerable interest.

In this paper we show that attosecond streaking opens the pathway towards a direct observation of matrix elements of the Eisenbud-Wigner-Smith (EWS) time delay operator \cite{Wig1955, Smi1960} down to a remarkable attosecond level. The EWS time delay is a measure for the spectral variation of the scattering phase. For later reference we note that the EWS time delay pertains to scattering in short-range potentials. Long-range Coulomb interactions are beyond the realm of standard EWS theory and require special care. As we will show, the extraction of the EWS time delays from streaking phases is highly non-trivial as the IR probe field and the long-range Coulomb field distort the spectral phase information to be extracted. The distortion can occur both in the entrance channel of the atomic bound state to be photoionized as well as in the exit channel of the continuum electronic wave packet leaving the ionic core. Remarkably, the latter can be accurately accounted for within a single-particle description by classical dynamics. A classical-trajectory Monte Carlo (CTMC) simulation is found to be in excellent agreement  with the numerical solution of the time-dependent Schr\"odinger equation (TDSE). For the entrance channel distortion we find a strong state and angular momentum dependence. In a proof-of-principle simulation we show that, after accounting for IR field distortion in the entrance and exit channels, the sensitivity of the streaking phase to short-ranged admixtures can reach the level of a few attoseconds.
Atomic units will be used throughout the text, unless indicated otherwise. \\

We numerically solve the full three-dimensional TDSE within a one-electron (or independent particle) framework,
\begin{multline}
\label{eq:TDSE}
i\frac{\partial}{\partial t} \Psi(\vec r, t) = \\
\left[-\frac{\vec \nabla^2}{2} + V(r) + \vec r\left(\vec F_{\mathrm{XUV}}(t)+ \vec F_{\mathrm{IR}}(t)\right) \right] \Psi(\vec r, t) \, ,
\end{multline}
where $\vec F_{\mathrm{XUV}}(t)$ is the (linearly polarized) electric field of the attosecond pump pulse ($\hbar\omega_{\mathrm{XUV}} \approx 100$\,eV), the intensity of which is sufficiently weak ($I_{\mathrm{XUV}} \lesssim 10^{13}$\,W/cm$^2$) as to allow for a perturbative treatment, and $\vec F_{\mathrm{IR}}(t)$ is the electric field of the few-cycle IR probe pulse ($\hbar\omega_{\mathrm{IR}} \approx 1.5$\,eV) with sufficient intensity in the non-perturbative regime such that the energy modulation by the streaking field provides energy (and time) resolution but remains below the level where tunnel ionization becomes prevalent ($I_{\mathrm{IR}} \approx 10^{10} - 10^{12}$\,W/cm$^2$). The electric fields are related to the corresponding vector potentials by $\vec F(t) = -\frac{\partial}{\partial t} \vec A(t)$. For the spherically symmetric atomic binding potential $V(r)$ we employ pure Coulomb potentials $V_C(r)=-\frac{Z}{r}$, as well as short-ranged model potentials.

Our computational method for solving \autoref{eq:TDSE} is based on the well-established pseudo-spectral split-operator method as described in \cite{TonChu1997}. The time-evolution is carried out using a split-operator algorithm where the evolution in the Coulomb field is performed in the energy representation of the wave function and the action of the external field in length gauge is calculated in coordinate space. The energy eigenstates are obtained by diagonalizing the field free Hamiltonian in a finite element discrete variable representation basis (cf.\ \cite{ResMcc2000,MccHorRes2001,SchCol2005,FeiNagPaz2008}). In order to disentangle channel distortion effects by the probe field from the ionization dynamics driven by the XUV field, we also consider a constrained TDSE (``restricted ionization model'' \cite{Schafer2009, MauJohMan2008}) where the wave function $\Psi(\vec r,t)$ is decomposed into the initial state $\Phi_0(\vec r)$ and the continuum part $\tilde \Phi(\vec r,t)$ created by absorption of a single photon from the XUV pulse. The IR field is constrained to act only on $\tilde \Phi(\vec r,t)$.

In our classical simulations we propagate electrons in the combined Coulomb and laser fields using a standard variable step-size Runge-Kutta algorithm. $10^7$ initial conditions were chosen either from the microcanonical ensemble for the binding energy of the level with principal quantum number $n$ or from its $(n,\ell,m)$ dependent subensembles \cite{YosReiBur1998} where $\ell$ and $m$ denote the orbital and magnetic quantum numbers, respectively. At the ionization time $\tau$ a randomly chosen energy from the energy spectrum of the XUV pulse (Gaussian distribution with a FWHM of 4\,eV) is transferred to the electron by adding a momentum in random direction. The ionization probability is taken to be proportional to the XUV intensity. \\

Starting point of the attosecond streaking technique \cite{ItaQueYud2002} is the assumption that the asymptotic momentum of a photoelectron ionized by an XUV pulse of duration of a few hundred attoseconds is shifted by the instantaneous value of the vector potential $\vec A_{\mathrm{IR}}(\tau)$ at the moment of ionization, i.e.\ arrival in the continuum,
\begin {equation}
\label{eq:streaking_simple}
\vec p_f(\tau)=\vec p_0 - \vec A_{\mathrm{IR}}(\tau)
\end{equation}
relative to the unperturbed asymptotic momentum of the photoelectron $p_0 = [2(\omega_{\mathrm{XUV}}-\mathcal{E}_i)]^\frac{1}{2}$, emitted by a photon with energy $\omega_{\mathrm{XUV}}$ and initial binding energy $\mathcal{E}_i$ (in a.u.).
For temporally well-defined IR electric streaking fields $\vec F_{\mathrm{IR}}(t)$ and vector potentials
\begin{equation}
\vec A_{\mathrm{IR}}(\tau) = - \int\limits_\tau^\infty  \vec F_{\mathrm{IR}}(t) dt
\end{equation}
the momentum shift and, hence, energy shift of the photoelectron allows thus, in principle, a mapping of emission time $\tau$ onto energy (``streaking'') and therefore the determination of emission time or, more precisely, of the phase $\phi_S = \omega_{\mathrm{IR}}\tau$ of a streaking spectrogram. An accuracy of the order of ten attoseconds or $\simeq\!0.05$ rad.\ can be reached for a streaking IR field with an oscillation period of the order of one femtosecond ($T \approx 2.6$\,fs for $\lambda_{\mathrm{IR}} = 800$ nm). The physical interpretation and analysis of streaking spectrograms and derived phases or times require, however, a detailed inquiry into the assumption underlying \autoref{eq:streaking_simple}.

The fundamental assumption of the streaking field as a probe is that the IR field does not distort the system to be probed. Since for easily resolvable energy shifts IR fields with intensities of the order $I_{\mathrm{IR}} \approx 10^{11} - 10^{12}$\,W/cm$^2$ are needed, distortion effects generally cannot be neglected. Unlike for the weak XUV pump pulse, effects beyond lowest-order perturbation theory must be considered for the IR field. As we will show in the following, the near independence of $\phi_S$ on $I_{\mathrm{IR}}$ within the accessible range does not contradict the notion that $\phi_S$ is strongly affected by the IR field. Moreover, the assumption of a ``sudden'' transition to the undisturbed momentum $p_0$ caused by photoabsorption (\autoref{eq:streaking_simple}) followed by the momentum shift by the IR field of the remaining streaking pulse requires scrutiny when we aim for the resolution of electronic dynamics on the attosecond scale. As the wave packet recedes from the ionic core, it propagates for $\sim 100$ as in the atomic (ionic) potential with a local momentum $\vec p(\vec r)$ rather than the asymptotic momentum $\vec p_0$. Neglect of the deviation of the local from the asymptotic momentum is in line with the strong-field approximation underlying the original attosecond streaking model \cite{ItaQueYud2002} but is no longer valid for streaking in the Coulomb field.

\begin{figure}[tb]
  \centering
  \includegraphics[width=\linewidth]{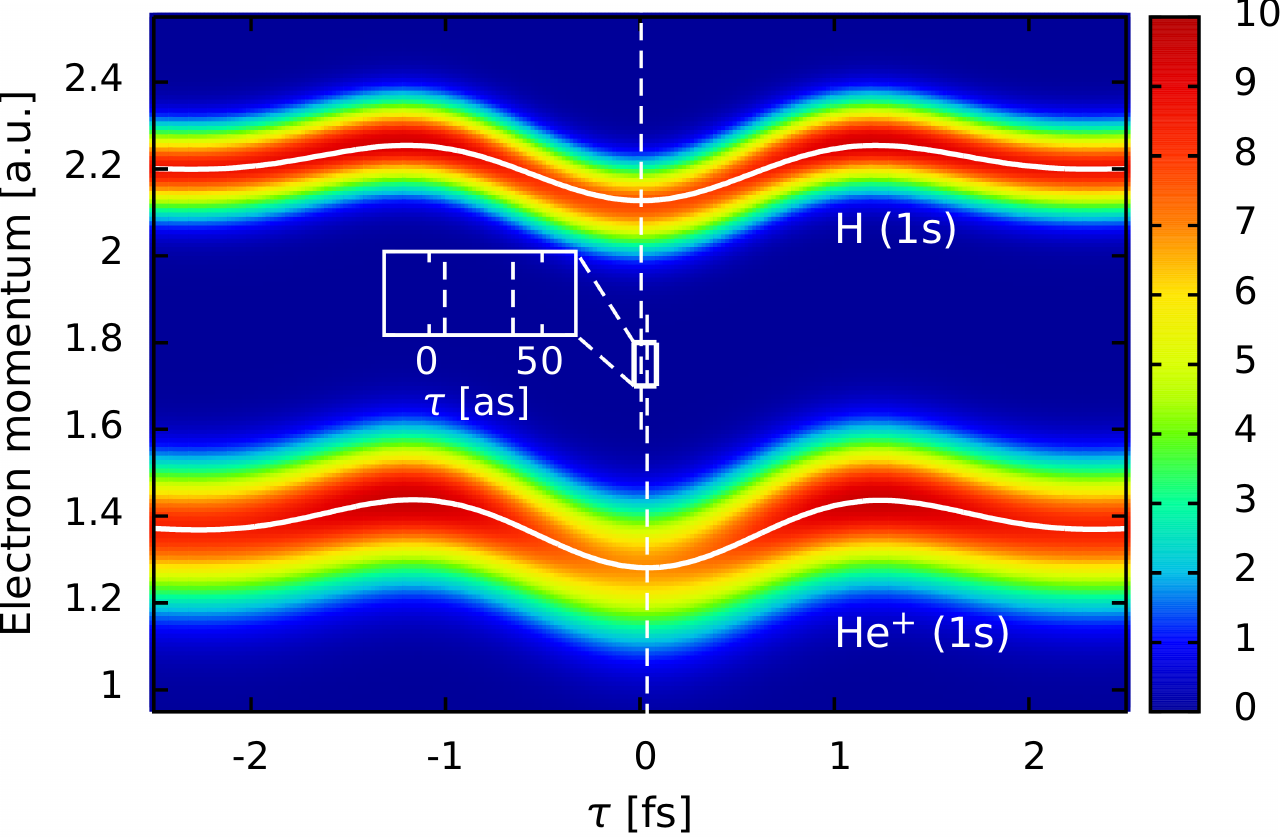} 
  \caption{Streaking spectrograms for an 800 nm IR laser field with a duration of 3\,fs and an intensity of $10^{12}$\,W/cm$^2$. The graphs show the final momentum distribution in ``forward'' direction of the laser polarization axis for H($1s$) and He$^+(1s)$ initial states and an XUV photon energy of 80\,eV. The solid white lines are the first moments of the electron spectra. The dashed white vertical lines indicate the shift of the central minimum relative to the vector potential. The colour scale is in arbitrary units.}
  \label{fig:spectrogram}
\end{figure}

Photoionization spectrograms for ionization of the H($1s$) and He$^+(1s)$ states by an 80\,eV XUV pulse with a duration of 200\,as (FWHM of the Gaussian intensity envelope), streaked by an 800\,nm IR laser field with a duration of 3\,fs (FWHM of the $\cos^2$ envelope) and an intensity of $10^{12}$\,W/cm$^2$ (as well as the first moment of the final electron momentum distribution, white lines) indicate the level of phase (or time) resolution achievable (\autoref{fig:spectrogram}).
By fitting the streaking curves to the analytic form of the IR vector potential $\vec A_{\mathrm{IR}}(t+t_S)$, we obtain an absolute time shift $t_S$ (with respect to the vector potential) of $-6.9$\,as for the ionization from the H($1s$) state and $-37.1$\,as for the ionization from the He$^+(1s)$ state, yielding a relative delay of $\Delta t_S = 30.2$\,as between the two emission channels. The sign convention for $t_S$ ensures that \emph{positive} values correspond to \emph{delayed} emission, i.e.\ the electron ``feels'' the vector potential at a \emph{later} time. The spectrograms are obtained with a full numerical solution of the 3D TDSE without any approximations. 
\begin{figure}[tb]
  \centering
  \includegraphics[width=\linewidth]{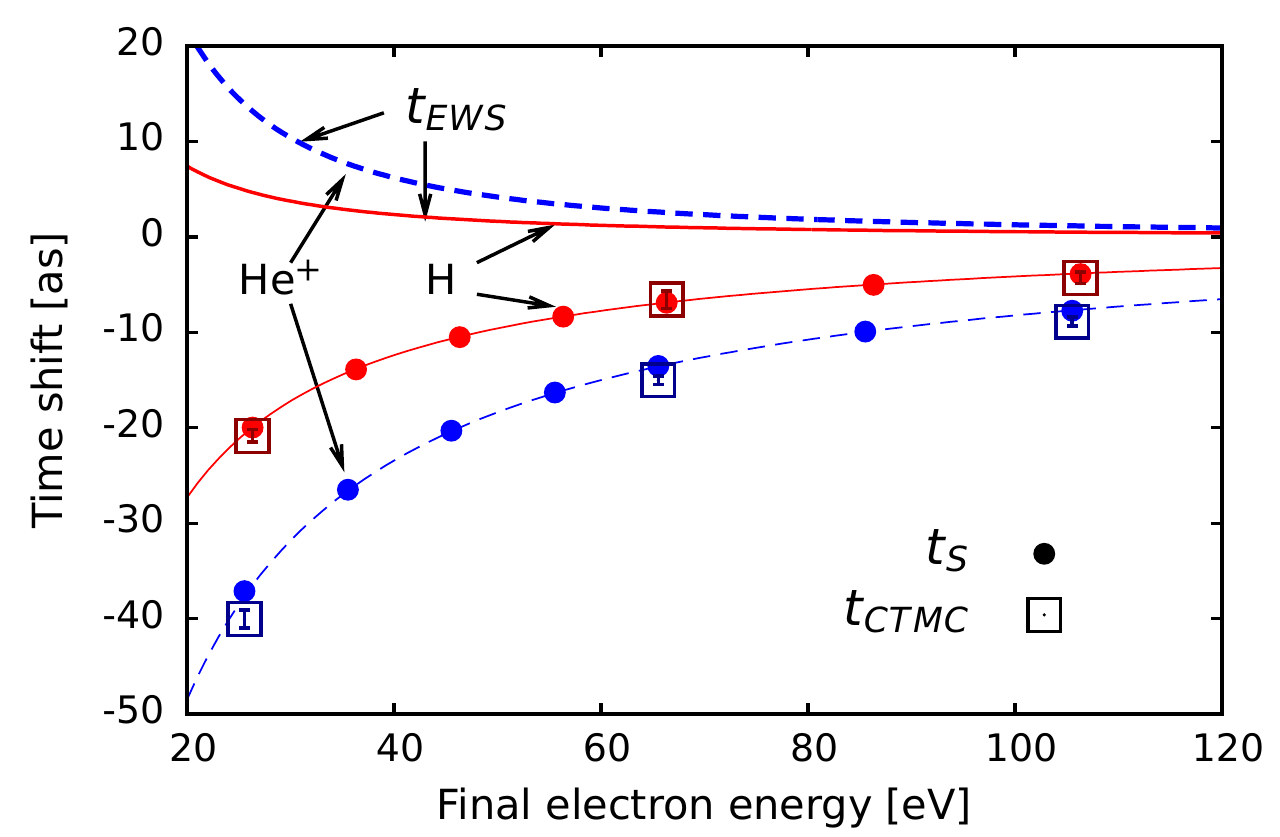}
  \caption{Temporal shifts $t_S$ extracted from quantum mechanical streaking simulations (H($1s$): red circles, He$^+(1s)$: blue circles), classical streaking simulations (H($1s$): red squares, He$^+(1s)$: blue squares), and for comparison, the EWS time shift $t_\mathrm{EWS} = d\varphi /dE$ applied to the Coulomb phase (H$^+$: red solid line, He$^{2+}$: blue dashed line).} 
  \label{fig:timeshifts}
\end{figure}
A negative time shift, or time advance, $t_S < 0$, relative to the vector potential can be observed over a wide range of final electron energies (\autoref{fig:timeshifts}) corresponding to varying the XUV pulse frequency $\omega_{\mathrm{XUV}}$. 
All results are obtained from spectrograms taken in ``forward'' direction with respect to the laser polarization axis and an opening angle of 10\,deg.\ (in the case of $s$-like initial states there are, however, no significant changes for the ``backward'' spectrograms and different opening angles). The error bars for the retrieved time shifts correspond to the estimated standard error from the nonlinear least-squares fit of the analytic vector potential to the calculated spectrograms.

The observed time shifts $t_S$ (or phase shifts $\phi_S$) are not directly related to the EWS-like time shift given by the energy derivative of the spectral phase $\varphi$, i.e.\ the group delay \cite{deCNus2002}
\begin{equation}
\label{eq:group_delay}
t_\mathrm{EWS} = \frac{d\varphi}{dE},
\end{equation}
also shown in \autoref{fig:timeshifts}. For the current case of photoionization from $1s$ initial states in Coulomb potentials the phase $\varphi$ would be given by the Coulomb phase
\begin{equation}
\label{eq:coulphase}
\varphi = \sigma_\ell = \arg[\Gamma(\ell+1-iZ/k)]
\end{equation}
where $\ell=1$ is the angular momentum of the free electron, $k$ its wavenumber, and $Z$ is the charge of the remaining ion. Even the sign of $t_\mathrm{EWS}$ predicted by Eqs.\ \ref{eq:group_delay} and \ref{eq:coulphase} is different from the streaking time shifts $t_S$ (\autoref{fig:timeshifts}). Obviously, $t_\mathrm{EWS}$ is overshadowed by the distortion of the outgoing Coulomb continuum electron by the IR field. It should be noted that the EWS time shift is only well-defined for short-ranged potentials. It is therefore no surprise that \autoref{eq:coulphase} is unrelated to what is observed in terms of the streaking phase.

Since classical-quantum correspondence holds for an unbound electron in both the Coulomb field (Rutherford scattering) and a laser field (Volkov states), it is suggestive to apply classical dynamics also to the motion of the outgoing electron in the combined Coulomb and IR fields in a streaking setting. Accordingly, the final momentum of the electron on its trajectory taking off at $\vec r(\tau) \approx 0$ near the ionic core is given by 
\begin{equation}
\label{eq:streaking_full}
\vec p_f(\tau) = \vec p_0 + \int\limits_{\tau}^\infty dt \, \vec a\left[\vec F_C(\vec r(t)), \vec F_\mathrm{IR}(t)\right] \, .
\end{equation}
In the limit of vanishing Coulomb field, $\vec F_C(t) = 0$, the acceleration $\vec a$ reduces to $\vec a[\vec F_\mathrm{IR}(t)] = \vec F_\mathrm{IR}(t)$ and \autoref{eq:streaking_full} becomes equal to \autoref{eq:streaking_simple}. For vanishing streaking field, $\vec F_\mathrm{IR}(t) = 0$, \autoref{eq:streaking_full} describes the approach of the instantaneous local momentum $\vec p(\tau)$ to the asymptotic momentum $\vec p_0$ along the Kepler hyperbola as $t \to \infty$ (or $r \to \infty$). The Coulomb corrected mapping from emission time to momentum shift thus becomes 
\begin{multline}
\label{eq:streaking_advanced}
 \vec p_f(\tau)=\vec p_0 - \vec A_\mathrm{IR}(\tau) + \\
\int \limits_{\tau}^{\infty} dt \, \left(\vec a\left[\vec F_C(\vec r(t)), \vec F_\mathrm{IR}(t)\right] - \vec a\left[\vec F_\mathrm{IR}(t)\right] \right) \, .
\end{multline}

\begin{figure}[tb]
  \centering
  \includegraphics[width=\linewidth]{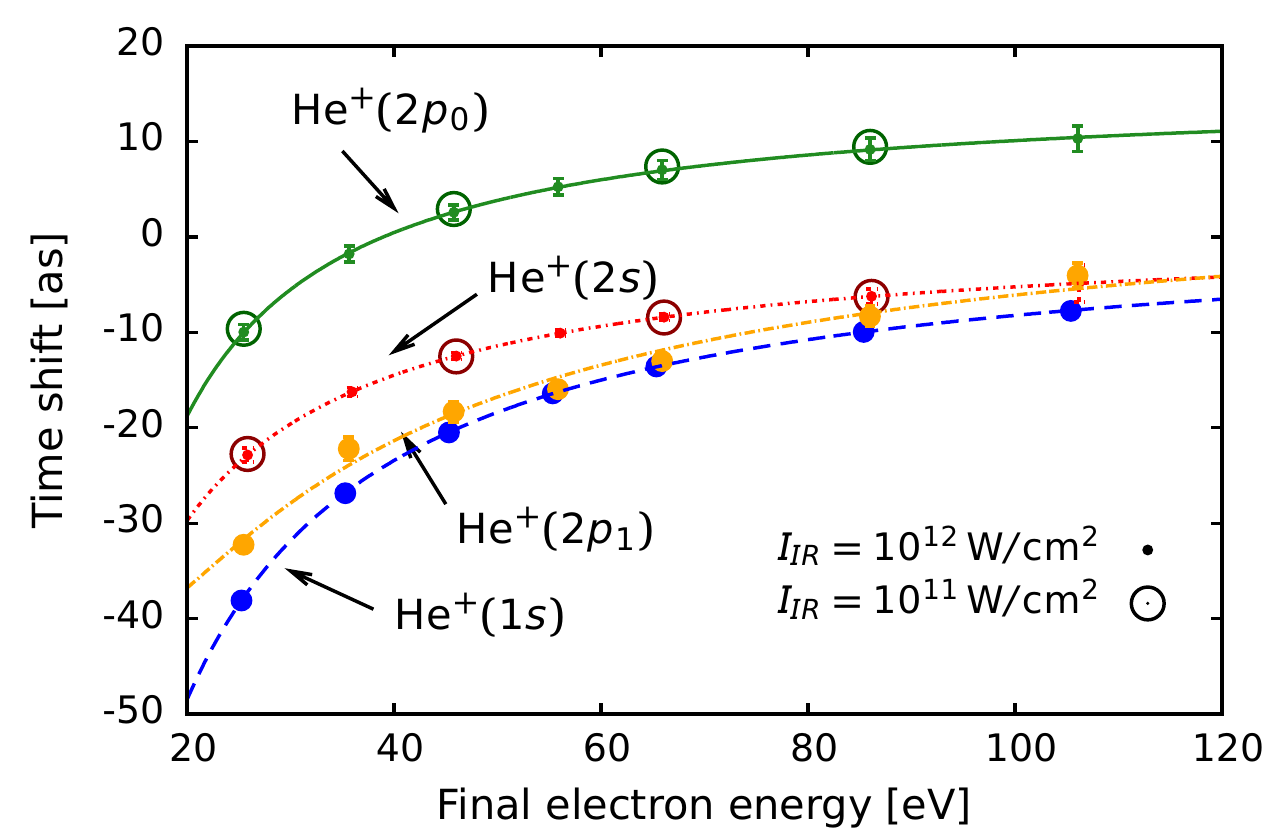}
  \caption{Streaking time shifts $t_S$ extracted from quantum mechanical streaking simulations for different initial states. The results are obtained from spectrograms taken in ``forward'' direction with respect to the laser polarization axis and an opening angle of 10\,deg.\,.}
  \label{fig:timeshifts_l}
\end{figure}

\autoref{eq:streaking_advanced} represents the classical realization of Coulomb-laser coupling (cf.\ \cite{SmiSpaIva2006a,SmiMouPat2007}) in the exit channel through the modification of the trajectory probing the Coulomb field by the simultaneous presence of the IR field. It treats the IR field and the Coulomb field non-perturbatively and on equal footing. One can easily show that the correction term relative to the standard streaking relation \autoref{eq:streaking_simple} yields, to leading order, a term $\sim\!\vec F_\mathrm{IR}(\tau)$ linear in the streaking field but phase shifted relative to $\vec A_\mathrm{IR}(\tau)$ by $\pi/2$. Consequently, \autoref{eq:streaking_full} becomes 
\begin{equation}
\label{eq:streaking_advanced_result}
\vec p_f(\tau) \approx \vec p_0 - \alpha\cdot\vec A_\mathrm{IR}(\tau + t_S) \, ,
\end{equation}
with $t_S$ the Coulomb-laser coupling induced time shift. Remarkably, $t_S$ is, to leading order, independent of the amplitude of the IR field since it depends on the interplay between $\vec F_\mathrm{IR}$ and $\vec A_\mathrm{IR}$ (cf.\,\autoref{fig:timeshifts_l}). 

Implementation of \autoref{eq:streaking_full} within our classical trajectory Monte Carlo (CTMC) simulation gives streaking time shifts in an almost perfect agreement with the full TDSE over a wide range of final-state energies for both H$(1s)$ and He$^+(1s)$ initial states (\autoref{fig:timeshifts}) and over two orders of magnitude in intensity ($I_\mathrm{IR} \approx 10^{10} - 10^{12}$\,W/cm$^2$) of the streaking field (not shown). The approximate independence of $t_S$ on the streaking field intensity following Eqs.\ \ref{eq:streaking_full} and \ref{eq:streaking_advanced_result} may mask the origin of $t_S$ to be, indeed, due to the Coulomb-laser coupling. We note that the present classical phase shift due to Coulomb-laser coupling appears to agree better with the full TDSE than the semi-classical estimates based on the eikonal approximation \cite{ZhaThu2010} which may be due to the perturbative approximation involved.

A second class of distortion effects refers to the entrance channel, i.e., to phase shifts of the initial state by the IR field. Clearly, those phase shifts are expected to be state dependent and are outside the scope of classical dynamics. We restrict our investigation to low-lying states of He$^+(n\ell m)$ which are stable against tunnel ionization by the IR field ($I_\mathrm{IR} \lesssim 10^{12}$\,W/cm$^2$). Within the full TDSE calculation we find remarkably strong initial-state dependent streaking field distortions. Most notably, the $2p$ substates aligned along ($m=0$) and perpendicular ($m=1$) to the polarization axis acquire a relative streaking phase shift of $\Delta t_S \approx 15$\,as, and the $2s$ relative to the $2p_0$ of $\Delta t_S \approx 10-20$\,as. This pronounced streaking phase shift as a function of angular momentum differs from recent calculations \cite{BagMad2010Err,ZhaThu2010} which did not find a shift for states with (initially) vanishing dipole moment. 

By performing constrained TDSE calculations (``restricted ionization model'') in which the initial-state interaction with the IR field is suppressed but continuum states fully evolve under the influence of the IR field, we find that the relative phase (time) shifts between $2s$, $2p_0$, and $2p_1$, for the most part, disappear. Thus, a state-dependent exit channel distortion can be ruled out. 
We find that also the initial-state dependent phase (time) shift is, within the range of useful streaking field intensities ($I_\mathrm{IR} \approx 10^{10} - 10^{12}$\,W/cm$^2$), remarkably insensitive to the value of $I_\mathrm{IR}$ (\autoref{fig:timeshifts_l}) closely following the behaviour observed for the exit channel distortion.
This apparent insensitivity to $I_\mathrm{IR}$ should not mislead to the conclusion that distortion effects by the IR field on the atomic dynamics can be excluded. 

The strong variation of $t_S$ with angular momentum $\ell$ is not specific to the degeneracy of the hydrogenic He$^+(n=2)$ manifold. By choosing an atomic model potential that breaks the $SO(4)$ symmetry we have verified that the streaking time shift persists for non-hydrogenic $\ell$ manifolds, thereby excluding degeneracy effects as the origin. Moreover, transient inter-shell coupling was found to dominate dynamical polarization over intra-shell mixing. One remarkable observation of the present results (\autoref{fig:timeshifts_l}) is that the relative time delay between an emitted $2p$ electron with 80\,eV final energy relative to a $2s$ electron with a final energy of 55\,eV yields $\Delta t_S \approx 20$\,as in striking agreement with the first experimental data for neon at the same final-state energies \cite{SchFieKar2010}. Clearly, this agreement may be, in fact, fortuitous as the electronic structure of neon is not properly accounted for. To the extent, however, that the dominant contribution to $\Delta t_S$ results from $\ell$ dependent entrance and quasi-classical exit channel IR distortions, the present results may, indeed, account for a considerable fraction of the delay observed. \\

So far, the present results quantitatively demonstrate the profound influence of the IR field on the observed streaking time shift. In the following, we will show that atomic photoemission times can be, indeed, accessed by streaking, provided that IR distortion effects can be either excluded or accounted for. We illustrate the power of attosecond streaking with the help of two prototypical examples.

\begin{figure}[tb]
  \centering
  \includegraphics[angle=0,width=\linewidth]{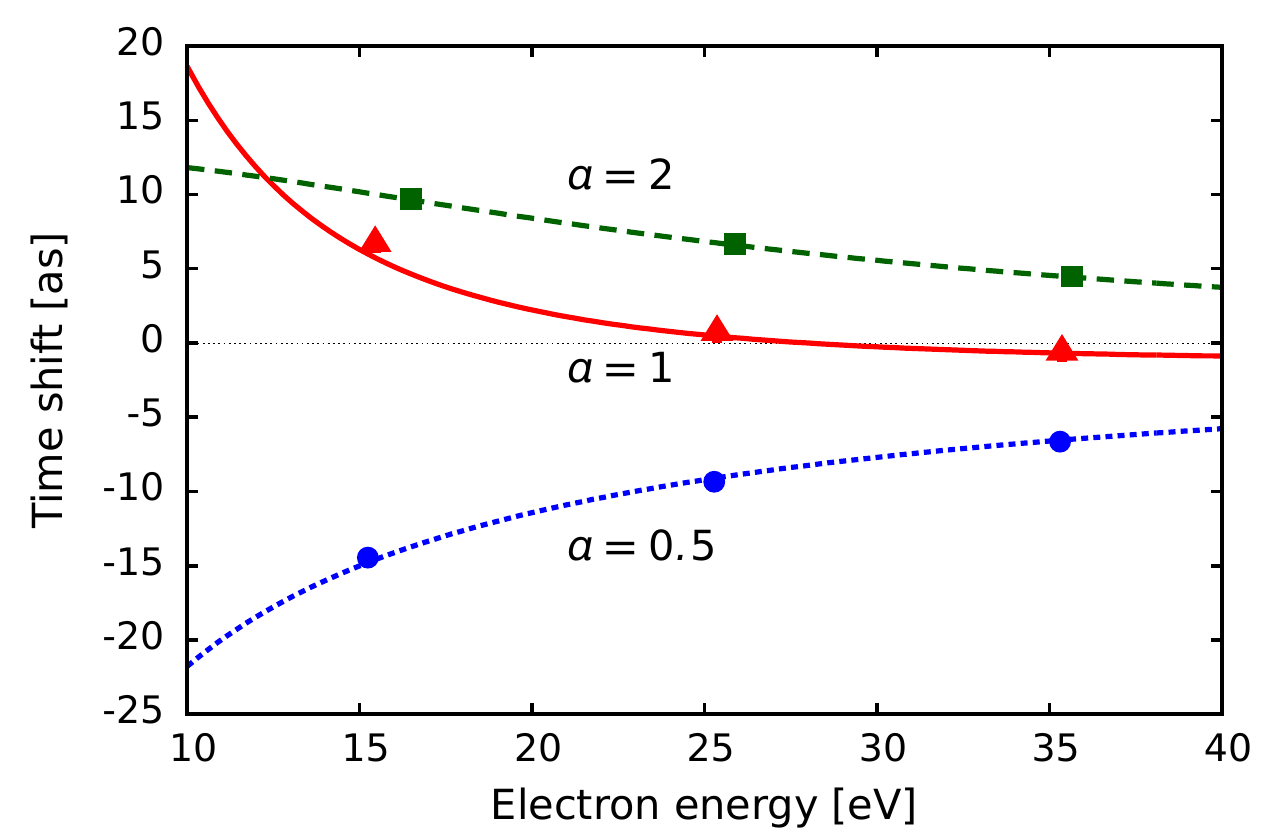}
  \caption{Temporal shifts $t_S$ extracted from quantum mechanical streaking simulations for the short-ranged Yukawa potential $V_\mathrm{Y}$ with screening lengths $a=0.5$ (squares), $a=1$ (triangles), and $a=2$ (diamonds) a.u.\,. The streaking IR laser field has a wavelength of 800\,nm, a duration of 3\,fs, and an intensity of $10^{12}$\,W/cm$^2$. Points: streaking shifts $t_S$, lines: $t_\mathrm{EWS}$ (Eisenbud-Wigner-Smith ``delays'').}
  \label{fig:timeshifts_Yukawa}
\end{figure}

We first consider an isotropic $1s$-like initial state deeply bound at $\mathcal{E}_i \approx -54.4$\,eV by a Yukawa-type short-ranged model potential with screening radius $a$,
\begin{equation}
V_\mathrm{Y}(r) = -\frac{Z}{r} e^{-\frac{r}{a}} \, .
\end{equation}
In this case, both initial-state dynamical polarization by the IR field and exit channel Coulomb-IR field coupling can safely be excluded. We calculate the EWS time delay or, equivalently, the energy variation of the spectral phase, $t_\mathrm{EWS} = d\varphi_\mathrm{Y}/dE$, for a wide range of final state energies and different combinations of charge $Z$ and atomic screening radii $a$ such that $\mathcal{E}_i$ remains constant. We find near perfect agreement with the corresponding time shift $t_S$ extracted from streaking spectrograms down to $\simeq\!15$\,eV final electron energy (\autoref{fig:timeshifts_Yukawa}).
Interestingly, $t_\mathrm{EWS}$ varies non-monotonically with $a$. Both time delays ($t_\mathrm{EWS} > 0$) and time advances ($t_\mathrm{EWS} < 0$) relative to the field $\vec A_\mathrm{IR}$ can occur. 
A physical realization of this scenario would be photodetachment of negative ions. This is, however, complicated by the fact that typical streaking field intensities would rapidly destroy the weakly bound negative-ion state.

As a second scenario we consider a short-ranged admixture to the Coulomb potential. In this case, the exit-channel distortion by the IR field is present but can be accurately accounted for (\autoref{eq:streaking_advanced}). For a short-ranged admixture of the form 
\begin{equation}
\label{eq:hump_potential}
\delta V(r) = V_0 \cdot e^{-\frac{(r-r_0)^2}{d^2}}
\end{equation}
to the tail of the Coulomb potential the residual EWS time shift can be determined. $\delta V(r)$ is chosen such that the initial state wave function and binding energy remain undistorted. This requires that $r_0 > \langle r\rangle_{1s}$ and $|\delta V(r)| \ll |V_C(r)|$ such that the photoelectron energy distribution remains undisturbed. We compare now the EWS time shift (averaged over the spectral width of the XUV pulse) induced by \autoref{eq:hump_potential} with the observed difference in the streaking time shift, $\Delta t_S = t_S - (t_S)_C$, after the Coulomb-laser coupling induced shift $(t_S)_C$ has been subtracted.
\begin{figure}[tb]
  \centering
  \includegraphics[angle=0,width=\linewidth]{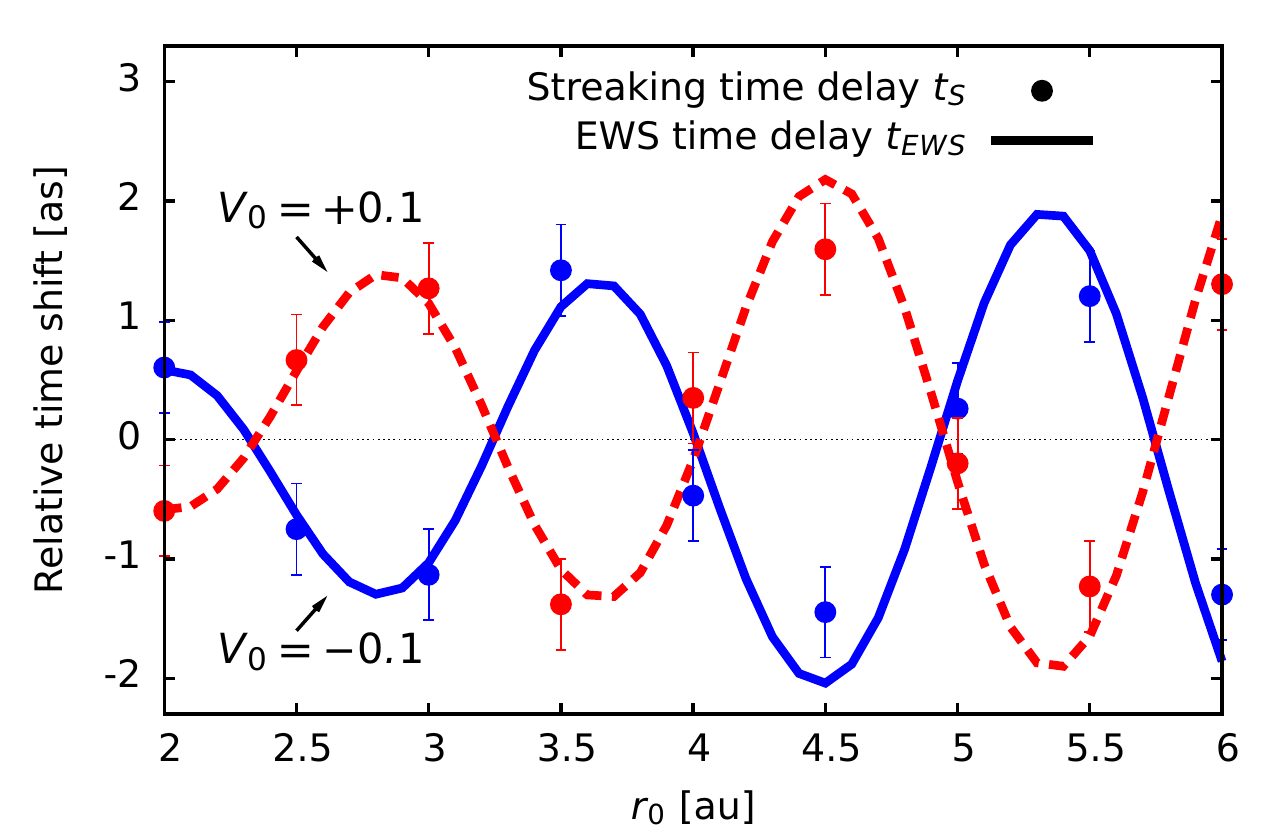}
  \caption{Streaking time shifts $\Delta t_S$ relative to the pure Coulomb case with $Z=2$ for a short-ranged admixture \autoref{eq:hump_potential} ($V_0 = \pm 0.1$\,a.u., $ d = 0.33$\,a.u.) at different positions $r_0$. The points are the time shifts as extracted from full streaking calculations for an XUV photon energy of $90$\,eV, an 800\,nm IR laser field with a duration of 3\,fs, and an intensity of $10^{12}$\,W/cm$^2$, the lines are $t_\mathrm{EWS}$ for the short-range admixture to the Coulomb potential, i.e.\ the Eisenbud-Wigner-Smith time ``delay''. The results of the EWS time delay were convoluted with the pure XUV spectrum to account for the spectral spread of the electronic wave packet.}
  \label{fig:coulomb_hump}
\end{figure}
We find remarkable agreement for different locations $r_0$ of the local perturbation and both signs of $V_0$ (\autoref{fig:coulomb_hump}). Even more remarkable is that the emission time shifts on the single-digit attosecond scale become accessible. A possible physical realization of this scenario would be photoemission from the central atom of endohedral $\mathrm{C}_{60}$ where the outgoing electron traverses the short-ranged shell potential of the cage. \\

In summary, we have identified on the one-electron (or independent particle) level considerable state-dependent time shifts that can be observed in attosecond streaking and which are of quantum mechanical origin. In addition, we have identified large time shifts which result from the coupling between the IR streaking field and the Coulomb field which depend on the final energy of the free electron and can be accounted for classically. The Eisenbud-Wigner-Smith (EWS) time shift (or energy variation of the scattering phase) is found to be accessible by streaking only if both initial-state dependent entrance channel and final-state exit channel distortions are properly accounted for. For such a scenario we have shown that time delays on the single-digit attosecond scale due to short-ranged potentials are in reach.

\ack
This work was supported by the FWF-Austria, Grants No.\ SFB016 and P21141-N16, the TeT under Grant No.\ AT-2/2009, and in part by the National Science Foundation through TeraGrid resources provided by NICS and TACC under Grant TG-PHY090031. The computational results presented have also been achieved in part using the Vienna Scientific Cluster (VSC). JF acknowledges support by the NSF through a grant to ITAMP.
The authors acknowledge valuable discussions with U.~Thumm and C.H.~Zhang on this subject.

\end{document}